\documentclass[11pt,a4paper]{article}
\usepackage{jheppub_kim}

\usepackage{pdflscape}
\usepackage{amsmath}
\usepackage{amssymb}
\usepackage{dcolumn}
\usepackage{bm}
\usepackage{color}
\usepackage{epsfig}
\usepackage{amsfonts}
\usepackage{graphicx}
\usepackage{subfigure}
\usepackage{dcolumn}

\newcommand{\be}{\begin{equation}}
\newcommand{\ee}{\end{equation}}
\newcommand{\bea}{\begin{eqnarray}}
\newcommand{\eea}{\end{eqnarray}}

\def\be{\begin{equation}}
\def\ee{\end{equation}}
\def\bea{\begin{eqnarray}}
\def\eea{\end{eqnarray}}

\begin{document}

\title{The imaginary potential and entropic force of heavy quarkonia in strongly coupled $ \mathcal{N}=4 $ supersymmetric Yang-Mills plasma on the Coulomb branch}
\author{M. Kioumarsipour}
\author{J. Sadeghi}

\affiliation{Sciences Faculty, Department of Physics, University of Mazandaran, Babolsar, Iran}

\emailAdd{m.kioumarsipour@stu.umz.ac.ir}

\abstract{\\
There are two important different mechanisms, the imaginary potential and entropic force, to investigate the dissociation of heavy quarkonia. In this paper, we calculate these two quantities for static and moving quarkonia in the rotating black 3-brane Type IIB supergravity solution dual to $ \mathcal{N}=4 $ super Yang-Mills theory on the Coulomb branch (cSYM) at strong coupling. At $ T\neq 0 $, there are two black hole branches: the large and small black hole branches. We investigate the effects of rotating parameter and rapidity for the static and moving quakonium at the large and small black hole branches. We find both mechanisms have the same results. In the large black hole branch: as $ T/\Lambda $ and $ \beta $ increase the thermal width decreases and so the suppression becomes stronger. In the small black hole branch: increasing $ T/\Lambda $ leads to increasing the thermal width and the quarkonium dissociates harder but $ \beta $ has an opposite effect.

\textbf{Keywords:} AdS/CFT correspondence, Quark-Gluon plasma, Imaginary potential, Entropic force.}

\maketitle

\section{Introduction}
In relativistic heavy-ion collisions done at the Relativistic Heavy Ion Collider (RHIC) and the LHC, a new phase of matter called quark-gluon plasma (QGP) has been produced \cite{q,g,p}. The heavy quarkonia can be created in the hard processes before the themalization of the plasma. They are useful probes to investigate the forming and evolving of the QGP
\cite{gg}. The AdS/CFT correspondence \cite{a,d,s} is a powerful and useful tool to investigate and compute the properties and hard probe parameters of a strongly coupled plasma. This correspondence connects the supergravity theories living in the $ d+1 $-dimensional AdS spacetimes to the quantum field theories living on the $ d $-dimension boundary.

One of the most significant experimental signatures of creating QGP is the melting of quarkonia say $J/\psi$ and excited states in the medium \cite{kar}. Heavy quark-antiquark potential $ V_{Q\overline{Q}} $ describes the interaction energy between a quark and an antiquark. At non zero temperature, the heavy quark potential may have an imaginary part which is related to the dissociation of the quarkonium. As argued in \cite{mat}, the color screening is the main mechanism responsible for this suppression. The imaginary part of the potential, $ImV_{Q\overline{Q}}$, is another more important reason than the color screening \cite{im1,im2}. The first calculation of the imaginary potential of quarkonium by using the AdS/CFT correspondence for $\mathcal{N}=4$ SYM theory was carried out by Noronha and Dumitru \cite{jon}. In fact, the imaginary potential is related to the thermal fluctuations stemmed from the interactions between the heavy quarks and the medium. Thereafter, people used this method several times, for example, for a static \cite{sta1,sta2} and moving quarkonium \cite{ali}, and finite 't Hooft coupling correction \cite{fad}. For more studies in different backgrounds, see \cite{ins,24,25,26,27,28}. In Refs. \cite{oth1,oth2}, other methods for studying $ImV_{Q\overline{Q}}$ have been introduced. 

Another important quantity that can be in charge of the suppression of the quarkonium is entropic force. The recent experimental research has shown the charmonium suppression at RHIC is stronger than at LHC in spite of its density is larger at LHC \cite{lh,lhc}. Kharzeev \cite{khar} has shown this conflict is related to the nature of deconfinement and the entropic force is responsible for dissociating the quarkonia. 
Hashimoto et al., calculating for the first time the entropic force associated with the heavy quark pair using the AdS/CFT correspondence \cite{hash}, showed that the entropy increases with the inter-quark distance and the peak of the entropy depends on the nature of deconfinement and emerges near the transition point. After those, the entropic force was calculated by using this method several times, for example, for a moving \cite{mov} and rotating quarkonium \cite{rot}, a moving dipole in a Yang-Mills like theory \cite{sara}, for the chemical potential effect \cite{che}, $ R^{2} $ and $ R^{4} $  corrections \cite{corr} and for the effect of deformation parameter \cite{def}.

The AdS/CFT correspondence was used to study the Coulomb branch of a strongly coupled $ \mathcal{N}=4 $ super Yang-Mills plasma. In this branch, the Higgs mechanism generates dynamically a scale $ \Lambda $. The scalar particles $ \Phi_{i}(i=1\cdots 6) $ of $ \mathcal{N}=4 $ SYM obtain a nonzero vacuum expectation value (VEV). This nonzero VEV breaks the conformal symmetry and the gauge symmetry $ SU(N_{c}) $ to its subgroup $ U(1)^{N_{c}-1} $, but does not break the supersymmetry and not result in a running of the coupling constant \cite{rr}. The thermodynamics and the hydrodynamic transport coefficients of $ \mathcal{N}=4 $ super Yang-Mills on the Coulomb branch are investigated in \cite{rot5,rot6}. Considering that the melting of a quarkonium is a significant experimental signature of creating QGP, in this paper we want to investigate two important mechanisms, the imaginary potential and the entropic force, on dual geometry of $ \mathcal{N}=4 $ SYM on the Coulomb branch (cSYM).

The paper is organized as follows. In section \ref{sec2}, we briefly review the rotating black $ 3 $-brane solution dual to $ \mathcal{N}=4 $ SYM on the Coulomb branch (cSYM) at strong coupling. In section \ref{sec3}, we investigate the imaginary part of potential a static and moving quarkonium in the rotating black $ 3 $-brane solution ($ \mathcal{N}=4 $ cSYM). In that section we consider two cases: the pair axis aligns perpendicularly and parallel to the plasma wind. In section \ref{ent}, we calculate the entropic force. In the last section, we summarize our results.

\section{Background geometry}\label{sec2}
The rotating black 3-brane solution found from the 5-dimensional Einstein-Maxwell-scalar action for the $ U(1)^{3} $ consistent truncation of Type IIB supergavity on $ S^{5} $ \cite{rot1,rot2,rot3,rot4,rot5,rot6}, is given by,
\begin{equation}
ds^{2}_{(5)}=\dfrac{r^{2}}{R^{2}}H^{1/3}\left (-fdt^{2}+dx_{1}^{2}+dx_{2}^{2}+dx_{3}^{2}\right )+\dfrac{H^{-2/3}}{\frac{r^{2}}{R^{2}}f}dr^{2}\label{1}
\end{equation}
where,
\begin{eqnarray}
f(r)&=&1-\dfrac{r_{h}^{4}}{r^{4}}\dfrac{H(r_{h})}{H(r)},~~~~~~H(r)=1-\dfrac{r_{0}^{2}}{r^{2}},\\
\varphi_{1}&=&\dfrac{1}{\sqrt{6}}\ln H, ~~~~~~~~~~~~~~~\varphi_{2}=\dfrac{1}{\sqrt{2}}\ln H,\nonumber\\
A_{t}^{1}&=&i\dfrac{r_{0}}{R^{2}}\dfrac{r_{h}^{2}\sqrt{H(r_{h})}}{r^{2}H(r)},\nonumber\\
r_{h}^{2}&=&\dfrac{1}{2}\left (r_{0}^{2}+\sqrt{r_{0}^{4}+4m}\right ),
\end{eqnarray}
where $ m $ is the mass parameter, $ A_{t}^{2}=A_{t}^{3}=0 $ and $ \kappa=\frac{r_{0}^{2}}{r_{h}^{2}} $. The metric (\ref{1}), after analytically continuing $ r_{0}\rightarrow -i\sqrt{q} $, is equivalent to the metric used in \cite{rot7}. An imaginary gauge potential does not lead to any inconsistencies in the bulk spacetime because all physical quantities in the 5-dimensional spacetime are given in terms of $ (\partial_{r}A^{1}_{t})^{2} $. Having an imaginary gauge potential or imaginary chemical potential $ \mu $, from the field theory side, means that the phase diagram of $ \mathcal{N}=4 $ cSYM is studied at finite temperature $ T $ and imaginary chemical potential $ \mu $. For the study of the QCD phase diagram on the lattice see \cite{ll1}.

The Hawking temperature $ T $ for the rotating black 3-brane solution (\ref{1}) is obtained as follows,
\begin{equation}
\dfrac{T}{\Lambda}=\dfrac{1-\frac{1}{2}\kappa}{\sqrt{\kappa-\kappa^{2}}},\label{z1}
\end{equation}
where $ \Lambda=\frac{r_{0}}{\pi R^{2}} $, $ T_{0}=\frac{r_{h}}{\pi R^{2}} $ and $ \kappa=\frac{r_{0}^{2}}{r_{h}^{2}}=\frac{\Lambda^{2}}{T_{0}^{2}} $. Also, one can invert (\ref{z1}) to obtain,
\begin{equation}
\kappa=\dfrac{1+\frac{T}{\Lambda}\left (\frac{T}{\Lambda}\mp \sqrt{\frac{T^{2}}{\Lambda^{2}}-2}\right )}{\frac{1}{2}+2\frac{T^{2}}{\Lambda^{2}}},\label{99}
\end{equation}
where ``-" and ``+" indicate the large and small black hole branches, respectively.

Now, we make the quarkonium moving. In order to make the moving pair, it is assumed that the frame is boosted in the $ x_{3} $ direction by rapidity $ \beta $ and the plasma is at rest. Then, by substituting,
\begin{eqnarray}
dt&=&cosh\beta dt'-sinh\beta dx'_{3},\nonumber\\
dx_{3}&=&-sinh\beta dt'+cosh\beta dx'_{3},\label{ll}
\end{eqnarray}
into (\ref{1}) and dropping the primes, we obtain the following boosted metric,
\begin{eqnarray}
ds^{2}&=&\dfrac{r^{2}}{R^{2}}H^{1/3}\Big [-\left (fcosh^{2}\beta -sinh^{2}\beta \right )dt^{2}+dx_{1}^{2}+dx_{2}^{2}\nonumber\\
&+&\big(cosh^{2}\beta -fsinh^{2}\beta \big)dx_{3}^{2}+2sinh\beta cosh\beta \big(f-1 \big)dtdx_{3} \Big]+\dfrac{R^{2}}{r^{2}fH^{2/3}}dr^{2}. \label{bb}
\end{eqnarray}
By setting $ \beta=0 $ in this equation, one will recover (\ref{1}). 

\section{Imaginary potential}\label{sec3}

In this section, we are going to investigate the imaginary potential of the quark-antiquark pair in $ \mathcal{N}=4 $ cSYM. In general, the quark-antiquark pair can orientate differently with respect to the plasma wind, i.e., transverse ($ \Theta=\pi/2 $) and parallel ($ \Theta=0 $).

\subsection{Transverse to the wind}
In the case of perpendicular to the wind in $ x_{1} $ direction, the parameterized coordinate is given by,
\begin{equation}
t=\tau,~~~~ x_{1}=\sigma,~~~~ x_{2}=0,~~~~x_{3}=0,~~~~r=r(\sigma),
\label{par} 
\end{equation}
where we consider the quark-antiquark pair situated at $ x_{1}=\pm L/2 $ and $ L $ denotes the inter-distance between the quark and antiquark.

To proceed, we begin with the Nambu-Goto action that is given by,
\begin{equation}
S=-\dfrac{1}{2\pi\alpha^{\prime}}\int d\tau d\sigma \mathcal{L} =-\dfrac{1}{2\pi\alpha^{\prime}}\int d\tau d\sigma \sqrt{-g},\label{NG}
\end{equation}
with $ g $ the determinant of the induced metric of the string worldsheet and,
\begin{equation}
g_{\alpha\beta}=G_{\mu\nu}\partial_{\alpha}X^{\mu} \partial_{\beta}X^{\nu},
\end{equation} 
where $ G_{\mu\nu} $ and $ X^{\mu} $ denote the metric and the target space coordinates, respectively.

By substituting (\ref{par}) into (\ref{bb}), one can obtain the induced metric as following,
\begin{eqnarray}
g_{00}&=&-\dfrac{r^{2}}{R^{2}}H^{1/3}\left(fcosh^{2}\beta -sinh^{2}\beta \right),\nonumber\\
g_{11}&=&\dfrac{r^{2}}{R^{2}}H^{1/3}\left( 1 +\dfrac{R^{4}r'^{2}}{r^{4}Hf} \right),~~~~~ r'=\frac{dr}{d\sigma},
\end{eqnarray}
and then the corresponding Lagrangian density becomes,
\begin{equation}
\mathcal{L}=\sqrt{a(r)+b(r)r'^{2}},\label{lag} 
\end{equation}
where,
\begin{eqnarray}
a(r)&=&\dfrac{r^{4}}{R^{4}}H^{2/3}\Big(fcosh^{2}\beta -sinh^{2}\beta \Big),\label{a(r)}\\
b(r)&=&H^{-1/3}\left(cosh^{2}\beta -\dfrac{sinh^{2}\beta}{f} \right).\label{b(r)}
\end{eqnarray} 
Since the Lagrangian (\ref{lag}) has no $ \sigma $-dependence explicitly, so the Hamiltonian $ H $ is a constant of motion,
\begin{equation}
H=\mathcal{L}-\dfrac{\partial \mathcal{L}}{\partial r^{\prime }}r^{\prime}=constant.
\end{equation}
By imposing the boundary condition, one can show the string configuration has a minimum at $ \sigma=0 $, i.e. $ r(0)=r_{c} $, so that $ r^{\prime}_{c}=0 $, and then,
\begin{equation}
r'=\sqrt{\dfrac{a^{2}(r)-a(r)a(r_{c})}{b(r)a(r_{c})}},\label{r'}
\end{equation}
with,
\begin{eqnarray}
a(r_{c})&=&\dfrac{r^{4}_{c}}{R^{4}}H(r_{c})^{2/3}\Big(f(r_{c})cosh^{2}\beta -sinh^{2}\beta \Big), \label{ac}\\
f(r_{c})&=&1-(\dfrac{r_{h}}{r_{c}})^{4}\dfrac{H(r_{h})}{H(r_{c})},~~~~~~H(r_{c})=1-(\dfrac{r_{0}}{r_{c}})^{2}.
\end{eqnarray}
For convenience, we set $ R=1 $. By integrating from (\ref{r'}), the inter-distance $ L $ of the quark-antiquark becomes,
\begin{equation}
L=2\int_{r_{c}}^{\infty}  dr \sqrt{\dfrac{b(r)a(r_{c})}{a^{2}(r)-a(r)a(r_{c})}}.\label{LL}
\end{equation}
Substituting (\ref{r'}) into (\ref{NG}), the action of the quark-antiquark pair reads,
\begin{equation}
S=\dfrac{\mathcal{T} }{\pi\alpha^{\prime}}\int_{r_{c}}^{\infty}  dr \sqrt{\dfrac{a(r)b(r)}{a(r)-a(r_{c}) }}. \label{11}
\end{equation}
The contributions of the free quark and antiquark cause divergency of this action. To avoid the divergency, one should subtract the self-energy of two quarks from (\ref{11}). The self-energy is given by,
\begin{equation}
S_{0}=\dfrac{\mathcal{T} }{\pi\alpha^{\prime}}\int_{r_{h}}^{\infty} dr \sqrt{b_{0}(r)}.
\end{equation}
where $ b_{0}(r)=\lim_{r\rightarrow \infty} b(r) $. Therefore, following the prescription in \cite{pott1,pott2,pott3} to study the real part of the potential, we have calculated the heavy quark potential as,
\begin{equation}
ReV_{Q\overline{Q}}=\dfrac{1}{\pi\alpha'}\int_{r_{c}}^{\infty} dr \left( \sqrt{\dfrac{a(r)b(r)}{a(r)-a(r_{c}) }}-\sqrt{b_{0}(r)}\right)-\dfrac{1}{\pi\alpha'}\int_{r_{h}}^{r_{c}}  dr \sqrt{b_{0}(r)}.
\end{equation}
\begin{figure}
\begin{center}$
\begin{array}{cc}
\includegraphics[width=73 mm]{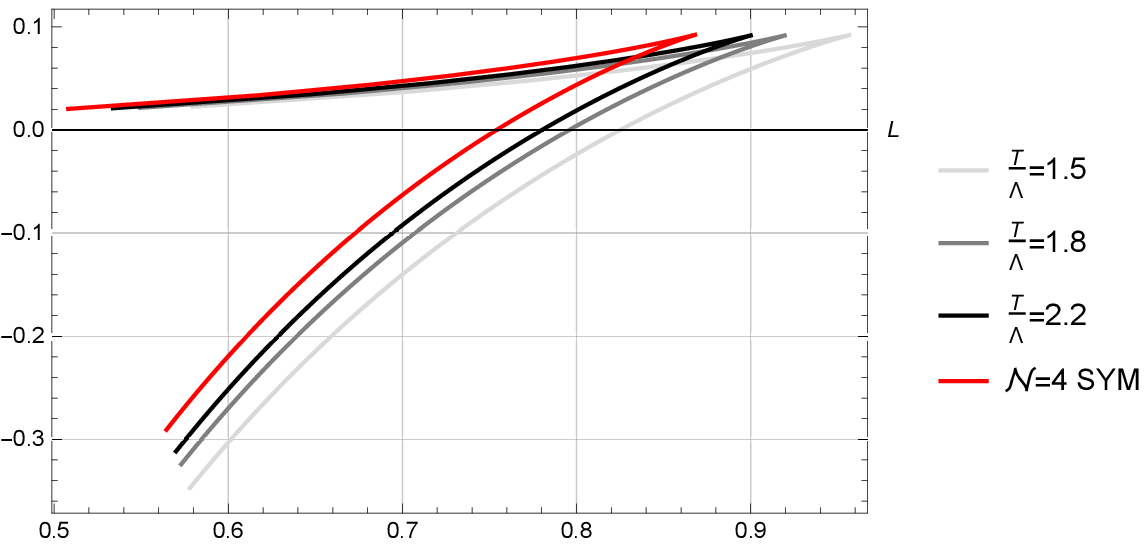}
\hspace{0.2cm}
\includegraphics[width=73 mm]{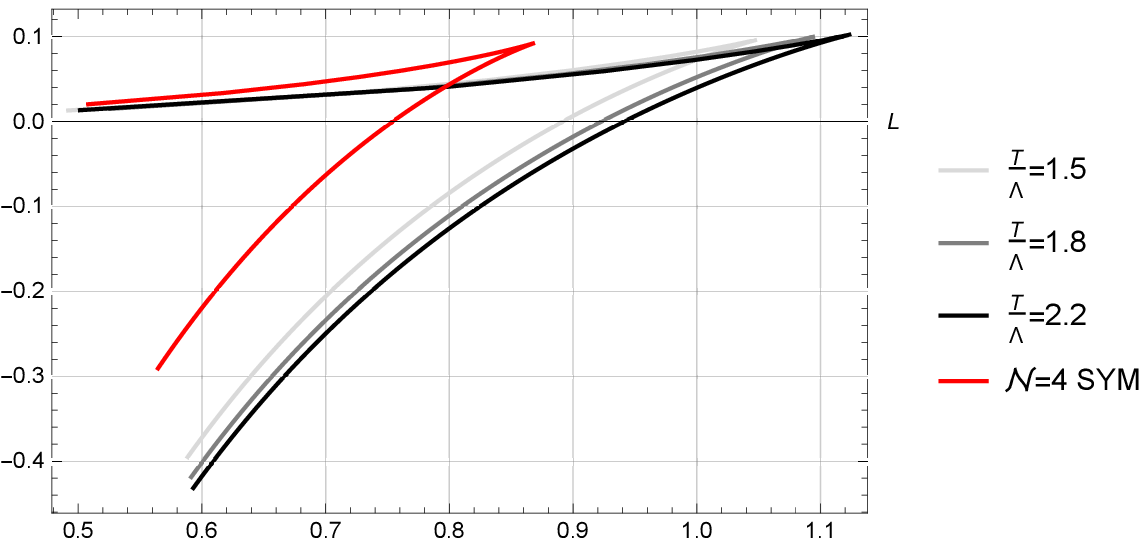}
\end{array}$
\end{center}
\caption{The heavy quark potential versus the interquark distance for different values $ T/\Lambda $: Left: Large blackhole, Right: Small blackhole.}
\label{vv}
\end{figure}

In order to see how $ T/\Lambda $ affects the static heavy quark potential, we plot numerically the potential $ V $ as a function of $ L $ at the fixed temperature ($ r_{h}=1 $) in figure \ref{vv}. As seen from the left panel, for the large black hole branch the potential increases as $ T/\Lambda $ increases and the dissociation length increases by decreasing $ T/\Lambda $. From the right panel, the small black hole branch, by increasing $ T/\Lambda $ the heavy potential decreases, and the dissociation length increases. The dissociation length in the small black hole is bigger than in the large black hole. Also, the quarkonium dissociates easier in $ \mathcal{N}=4 $ SYM than in both the large and small black holes.

On the other hand, by applying the thermal worldsheet fluctuation method \cite{jon} for studying the imaginary part of the heavy quark potential we obtain,
\begin{equation}
ImV_{Q\overline{Q}}=-\dfrac{1}{2\sqrt{2}\alpha'}\sqrt{b(r_{c})}\left(\dfrac{a'(r_{c})}{2a''(r_{c})}-\dfrac{a(r_{c})}{a'(r_{c})}\right),\label{imm} 
\end{equation}
where,
\begin{eqnarray}
a'(r_{c})&=&r_{c}^{4}H(r_{c})^{2/3} \left[f'({r_{c}})cosh^{2}\beta +\left (\dfrac{4}{r}+\dfrac{2}{3}\dfrac{H'(r_{c})}{H(r_{c})}\right ) \left (f(r_{c})cosh^{2}\beta -sinh^{2}\beta\right )\right ],\\
a''(r_{c})&=&r_{c}^{4}H(r_{c})^{2/3} \Bigg[ f''({r_{c}})cosh^{2}\beta +2\left (\dfrac{4}{r}+\dfrac{2}{3}\dfrac{H'(r_{c})}{H(r_{c})}\right ) f'(r_{c})cosh^{2}\beta \\
&+&\!\!\left  (\left  (\dfrac{4}{r}+\dfrac{2}{3}\dfrac{H'(r_{c})}{H(r_{c})}\right  )^{2} +\dfrac{2}{3}\dfrac{H''(r_{c})H(r_{c})-H'(r_{c})^{2}}{H(r_{c})^{2}}-\dfrac{4}{r_{c}^{2}}\right  ) \left (f(r_{c})cosh^{2}\beta -sinh^{2}\beta\right )\!\Bigg],\nonumber\\
b(r_{c})&=&H(r_{c})^{-1/3}\left[cosh^{2}\beta -\dfrac{sinh^{2}\beta}{f(r_{c})} \right],\label{bcc}
\end{eqnarray}
with,
\begin{eqnarray}
H'(r_{c})&=&\dfrac{2r_{0}^{2}}{r_{c}^{3}},~~~~~~H''(r_{c})=\dfrac{-6r_{0}^{2}}{r_{c}^{4}},\\
f'(r_{c})&=&\dfrac{r_{h}^{4}}{r_{c}^{5}}\left (\dfrac{1-(r_{0}/r_{h})^{2}}{\big [1-(r_{0}/r_{c})^{2}\big]^{2}}\right )\left (2-\left (\dfrac{r_{0}}{r_{c}}\right )^{2}\right ),\label{ffc}\\ 
f''(r_{c})&=&-\dfrac{r_{h}^{4}}{r_{c}^{6}}\left (\dfrac{1-(r_{0}/r_{h})^{2}}{\big [1-(r_{0}/r_{c})^{2}\big ]^{3}}\right )\left (6\left (\dfrac{r_{0}}{r_{c}}\right )^{4}-18\left  (\dfrac{r_{0}}{r_{c}}\right )^{2}+20\right ),\label{fff}
\end{eqnarray}
where the derivatives are with respect to $ r $. Also, note if $ r_{0}=0 $, the results of \cite{ali} are reproduced, and if $ r_{0}=\beta=0 $ the results of $ \mathcal{N}=4 $ SYM plasma \cite{jon} are recovered.

Now, as mentioned in \cite{jon,fi}, we should discuss three restrictions on the formula (\ref{imm}). First, the term $ b(r_{c}) $ must be positive, then,
\begin{equation}
\xi < \xi_{max}=\sqrt{\dfrac{1-tanh^{2}\beta}{2(1-\kappa)}\left (-\kappa+\sqrt{\kappa^{2}+\dfrac{4(1-\kappa)}{1-tanh^{2}\beta}}\right )}
\end{equation}
where $ \xi\equiv r_{h}/r_{c} $.

Second, since the imaginary part of potential should be negative, so,
\begin{equation}
\dfrac{a'(r_{c})}{2a''(r_{c})}-\dfrac{a(r_{c})}{a'(r_{c})} >0,
\end{equation}
results in,
\begin{equation}
\xi_{min} <\xi,
\end{equation}
and the value of $ \xi_{min} $ can be evaluated numerically.

The third limitation pertains to the maximum value of the $ LT(\xi) $, $ LT_{max} $, showing the limit of validity of the saddle point approximation. To address this, we plot $ LT $ as a function of $ \xi $ for different values of $ T/\Lambda $ and $ \beta $ numerically. As seen clearly, in each plot each line has a maximum value of $ LT $. $ LT $ ascends as $ \xi $ increases up to the critical value $ \xi_{max} $ and descends for $ \xi_{max}<\xi $. In fact, for very large distances (the region of $ \xi_{max}<\xi $) one should consider highly curved configurations that are not solutions of the Nambu-Goto action \cite{bac}. Hence, we will consider only the region of $ \xi<\xi_{max}  $. 

\begin{figure}
\begin{center}$
\begin{array}{cc}
\includegraphics[width=73 mm]{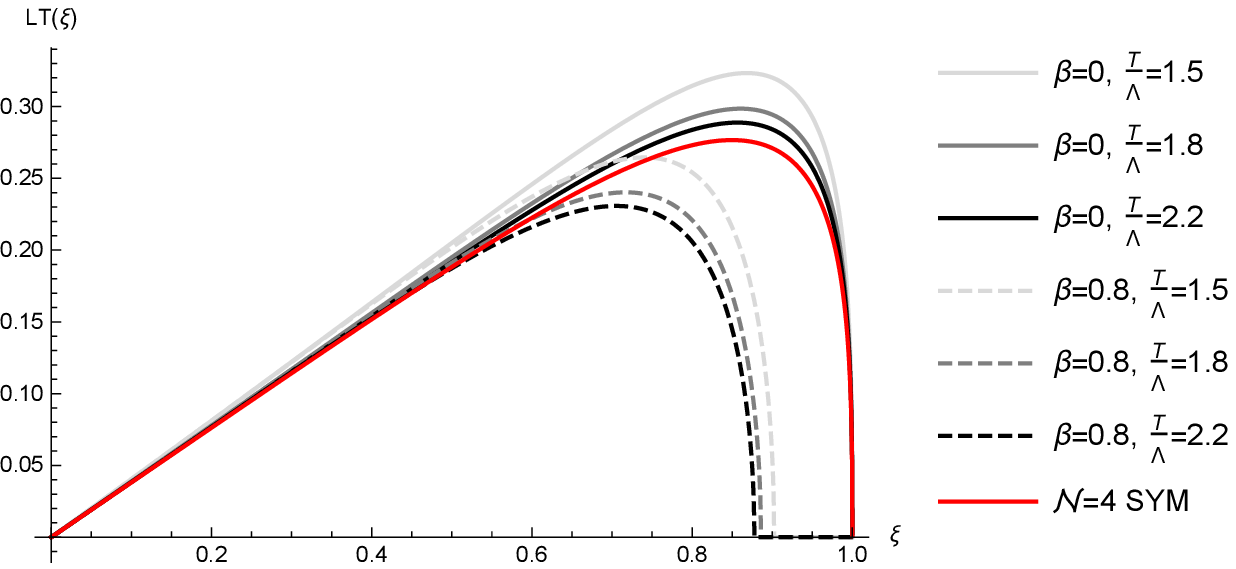}
\hspace{0.2cm}
\includegraphics[width=73 mm]{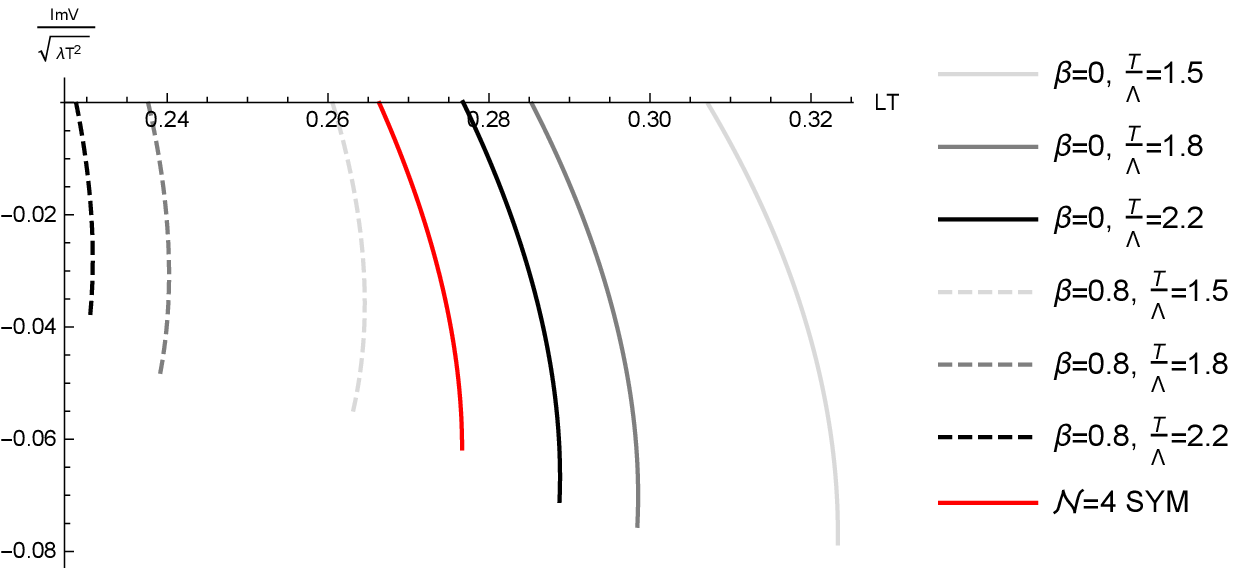}
\end{array}$
\end{center}
\caption{The large black hole. Left: plot of $ LT $ versus $ \xi $ for different values of $ T/\Lambda $ and $ \beta $. Right: $ ImV_{Q\overline{Q}}/\sqrt{\lambda}T $ versus $ LT $ for different values of $ T/\Lambda $ and $ \beta $.}
\label{gg}
\end{figure}

\begin{figure}
\begin{center}$
\begin{array}{cc}
\includegraphics[width=73 mm]{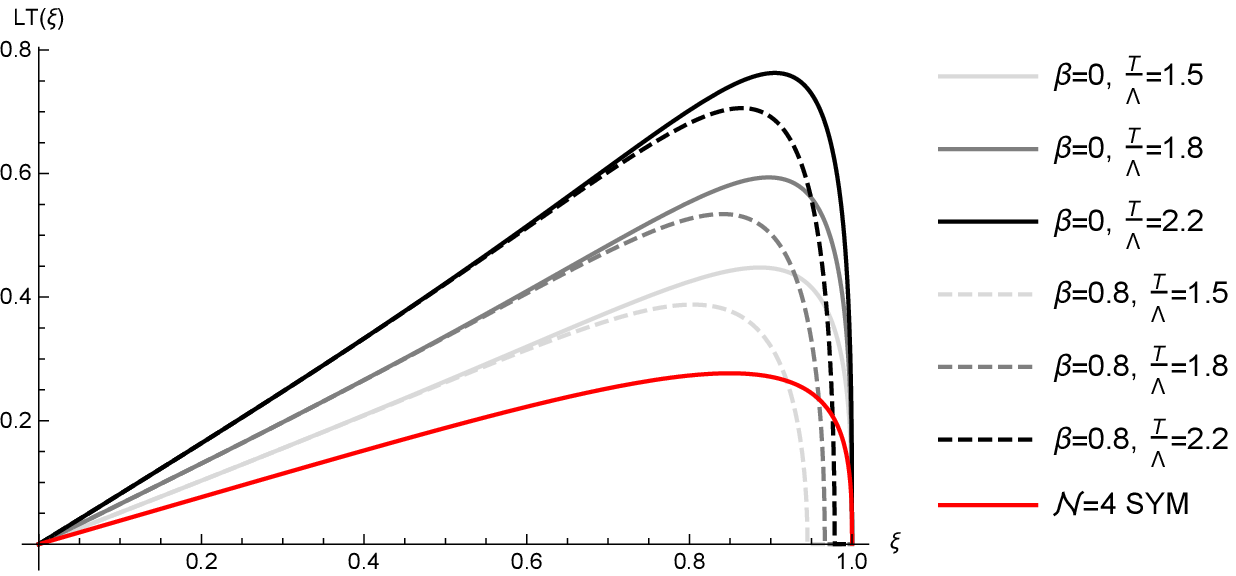}
\hspace{0.2cm}
\includegraphics[width=73 mm]{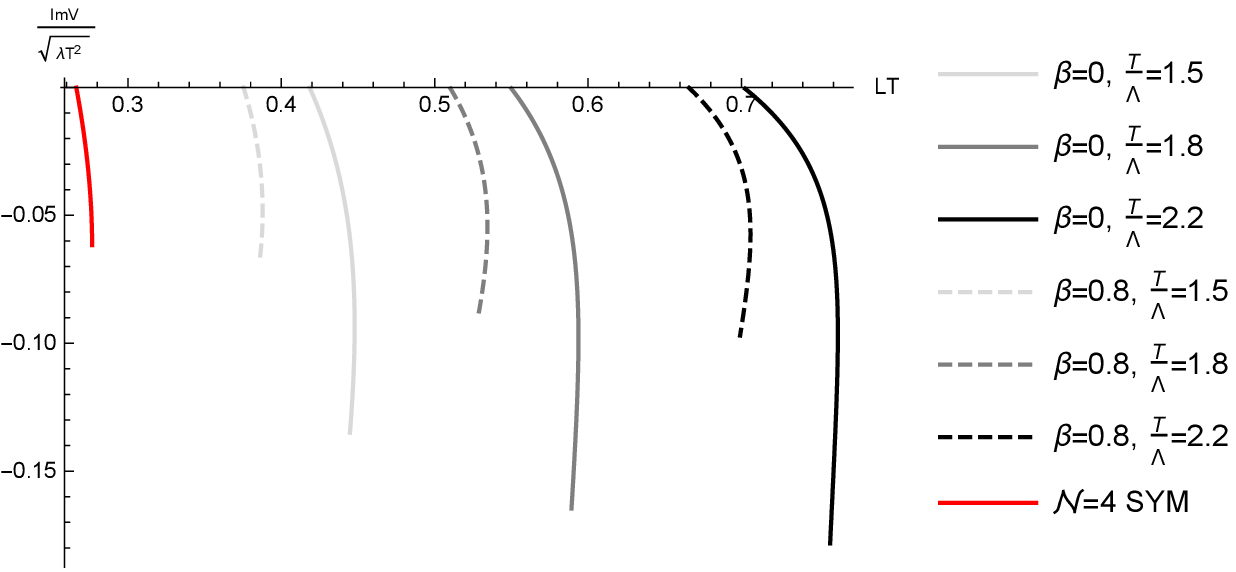}
\end{array}$
\end{center}
\caption{The small black hole. Left: plot of $ LT $ versus $ \xi $ for different values of $ T/\Lambda $ and $ \beta $. Right: $ ImV_{Q\overline{Q}}/\sqrt{\lambda}T $ versus $ LT $ for different values of $ T/\Lambda $ and $ \beta $.}
\label{gggg}
\end{figure}

Here, we investigate the effects of  $ T/\Lambda $ and rapidity $ \beta $ on the inter-distance and the imaginary potential. In figures \ref{gg} and \ref{gggg}, we plot numerically the $ LT $ as a function of $ \xi $, and $ ImV_{Q\overline{Q}}/(\sqrt{\lambda}T) $ versus $ LT $ (at different values of $ T/\Lambda $ and $ \beta $) for the large and small black holes, respectively. 

\begin{itemize}
\item The large black hole\\
 As seen in the left panel figure \ref{gg}, increasing $ T/\Lambda $ leads to decreasing $ LT_{max} $ at the fixed rapidity $ \beta $. Also, $ LT_{max} $ decreases as $\beta$ increases at the fixed $ T/\Lambda $. One sees that increase of $ T/\Lambda $ and $ \beta $ lead to decrease $ LT_{max} $. A corresponding value of $ LT $ for the case of $ \mathcal{N}=4 $ SYM is $ LT_{max}=LT(\xi_{max})\sim 0.28 $ \cite{sta1}. The corresponding values of $ LT $ are bigger and less than in $ \mathcal{N}=4 $ SYM for the static quarkonium and moving quarkonium, respectively.
We can see from the right panel that the imaginary potential starts at a $ L_{min} $ which is solution of $ ImV_{Q\overline{Q}}=0 $, and ends at a $ L_{max} $. The absolute value of the imaginary potential decreases by increasing $ T/\Lambda $ at the fixed $ \beta $. In fact, $ T/\Lambda $ generates the $ ImV_{Q\overline{Q}} $ for smaller inter-distance. As we know, if the onset of the $ ImV_{Q\overline{Q}}$ happens for smaller $LT$ the suppression will be stronger \cite{fi}. Therefore, $ T/\Lambda $ tends to decrease the thermal width and makes the suppression stronger for the large black hole. Also, as the rapidity $ \beta $ increases, the absolute value of the imaginary potential decreases and the suppression becomes stronger at the fixed $ T/\Lambda $; this result is consistent with \cite{ali,fad}. Also, we can see the static quarkonium dissociates harder in $ \mathcal{N}=4 $ cSYM than in $ \mathcal{N}=4 $ SYM. By increasing rapidity $ \beta $, the imaginary potential happens for smaller $ LT $, therefore the moving quarkonium can dissociate easier in $ \mathcal{N}=4 $ cSYM than in $ \mathcal{N}=4 $ SYM.
\item The small black hole\\
From the left panel figure \ref{gggg}, one can see that the values of $ LT_{max} $ increase as $ T/\Lambda $ increases at the fixed $ \beta $ and decrease as $ \beta $ increases. In the small black hole branch, the values of $ LT_{max} $ are bigger than in $ \mathcal{N}=4 $ SYM. As seen from the right panel, the absolute value of the imaginary potential increases by increasing $ T/\Lambda $ at the fixed $ \beta $. In fact, $ T/\Lambda $ generates the $ ImV_{Q\overline{Q}} $ for larger inter-distance, increases the thermal width, and makes the suppression weaker for the small black hole. Increasing $ \beta $ at the fixed $ T/\Lambda $ leads to decreasing the thermal width and makes the suppression stronger. Also, one can see that the quarkonium dissociates easier in $ \mathcal{N}=4 $ SYM than in the small black hole branch.
\end{itemize}


\subsection{Parallel  to the wind}
In following, we consider the system oriented parallel to the wind in $ x_{3} $ direction,
\begin{equation}
t=\tau,~~~~ x_{1}=0,~~~~x_{2}=0,~~~~x_{3}=\sigma,~~~~ r=r(\sigma),
\end{equation}
where the quark and antiquark are located at $ x_{3}=\pm L/2 $.

By performing the similar calculation in the previous subsection, we obtain the following inter-distance,
\begin{equation}
L=2\int_{r_{c}}^{\infty} dr \sqrt{\dfrac{A(r_{c})B(r)}{A^{2}(r)-A(r)A(r_{c})}},
\end{equation}
where,
\begin{eqnarray}
A(r)=\dfrac{r^{4}}{R^{4}}H(r)^{2/3}f(r),&&~~~~~~A(r_{c})=\dfrac{r_{c}^{4}}{R^{4}}H(r_{c})^{2/3}f(r_{c}),\nonumber\\ \label{AR}
B(r)=H^{-1/3}&&\left(cosh^{2}\beta -\dfrac{sinh^{2}\beta}{f} \right).
\end{eqnarray}
The imaginary part of the potential for the parallel case is given by,
\begin{equation}
ImV_{Q\overline{Q}}=-\dfrac{1}{2\sqrt{2}\alpha'}\sqrt{B(r_{c})}\left(\dfrac{A'(r_{c})}{2A''(r_{c})}-\dfrac{A(r_{c})}{A'(r_{c})}\right),\label{im} 
\end{equation}
where,
\begin{eqnarray}
A'(r_{c})&=&\dfrac{r_{c}^{4}H(r_{c})^{2/3}}{R^{4}}\left [f'(r_{c})+f(r_{c})\left (\dfrac{4}{r_{c}}+\dfrac{2}{3}\dfrac{H'(r_{c})}{H(r_{c})}\right )\right ],\label{aaa} \\
A''(r_{c})&=&\dfrac{r_{c}^{4}H(r_{c})^{2/3}}{R^{4}}\Bigg [f''(r_{c})+2f'(r_{c})\left (\dfrac{4}{r_{c}}+\dfrac{2}{3}\dfrac{H'(r_{c})}{H(r_{c})}\right )\nonumber\\
&+&\left ( \left (\dfrac{4}{r_{c}}+\dfrac{2}{3}\dfrac{H'(r_{c})}{H(r_{c})}\right )^{2}+\dfrac{2}{3}\dfrac{H''(r_{c})H(r_{c})-H'(r_{c})^{2}}{H(r_{c})^{2}}-\dfrac{4}{r_{c}^{2}}\right )f(r_{c})\Bigg ],~~~~~~~\\
B(r_{c})&=&H(r_{c})^{-1/3}\left[cosh^{2}\beta -\dfrac{sinh^{2}\beta}{f(r_{c})} \right],
\end{eqnarray}
and $ f'_{c} $ and $ f''_{c} $ are defined in (\ref{ffc}) and (\ref{fff}), respectively.

\begin{figure}
\begin{center}$
\begin{array}{cc}
\includegraphics[width=73 mm]{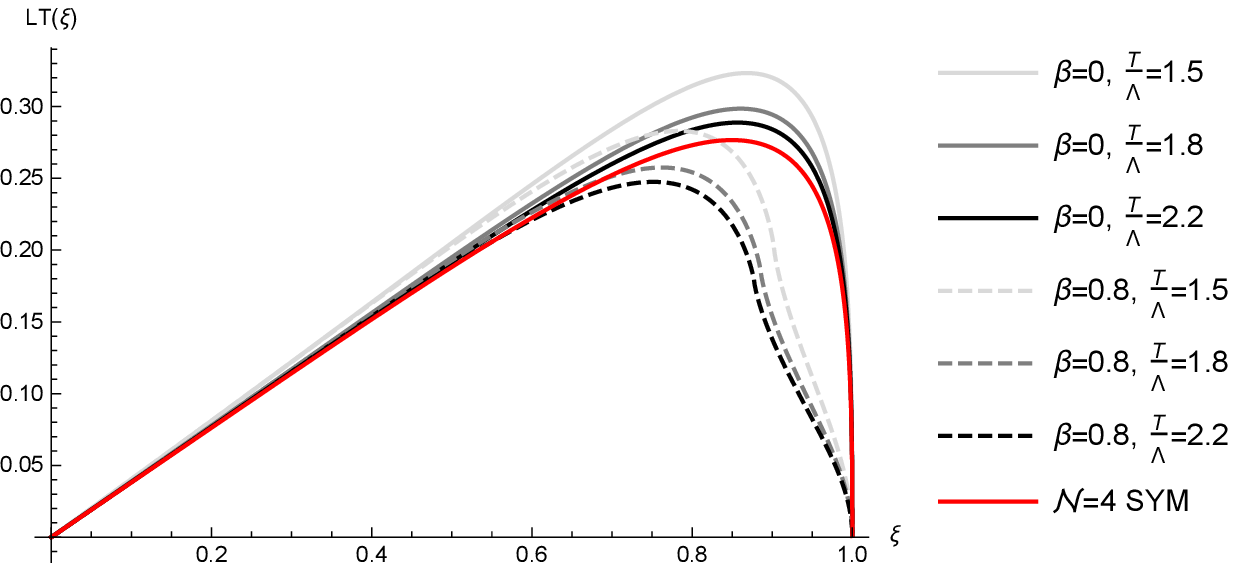}
\hspace{0.2cm}
\includegraphics[width=73 mm]{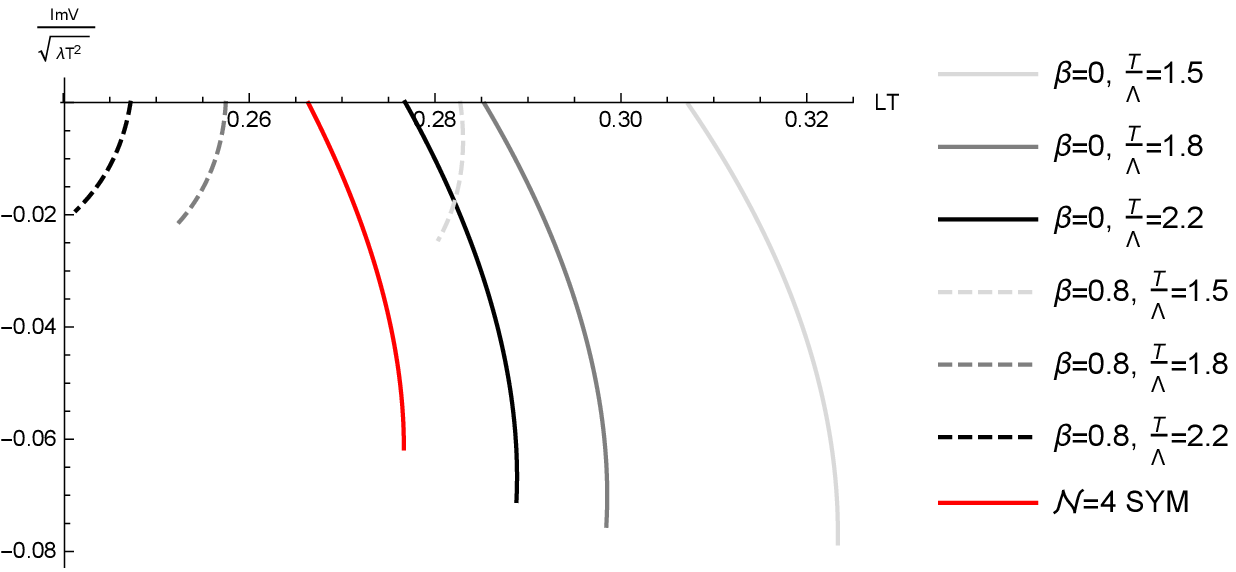}
\end{array}$
\end{center}
\caption{The large black hole. Left: plot of $ LT $ versus $ \xi $ for different values of $ T/\Lambda $ and $ \beta $. Right: $ ImV_{Q\overline{Q}}/\sqrt{\lambda}T $ versus $ LT $ for different values of $ T/\Lambda $ and $ \beta $.}
\label{w2}
\end{figure}

\begin{figure}
\begin{center}$
\begin{array}{cc}
\includegraphics[width=73 mm]{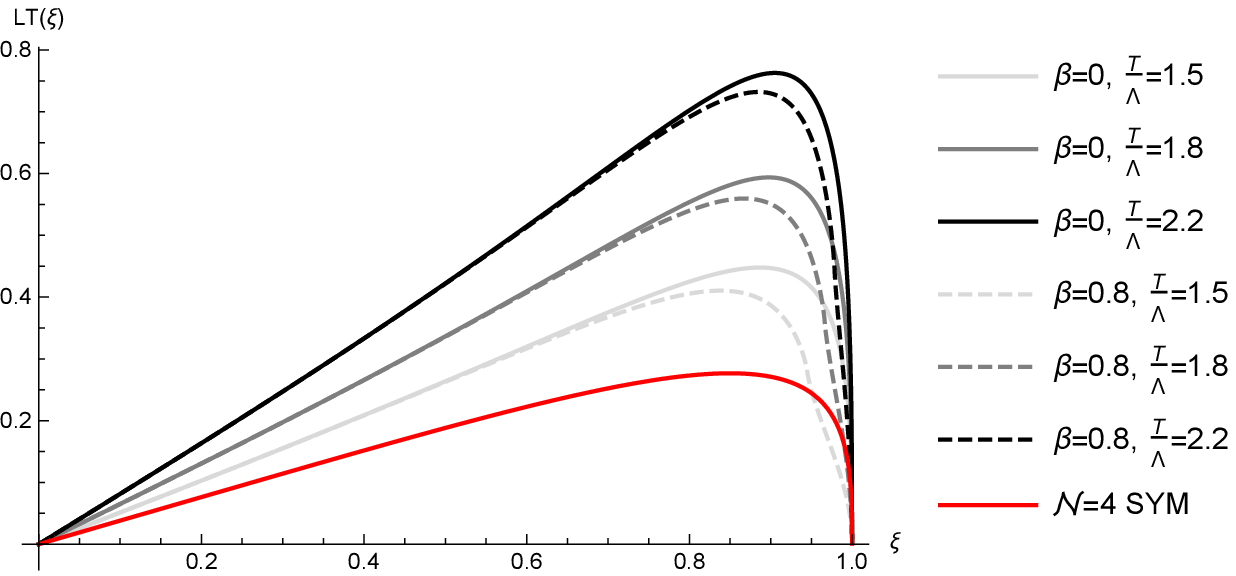}
\hspace{0.2cm}
\includegraphics[width=73 mm]{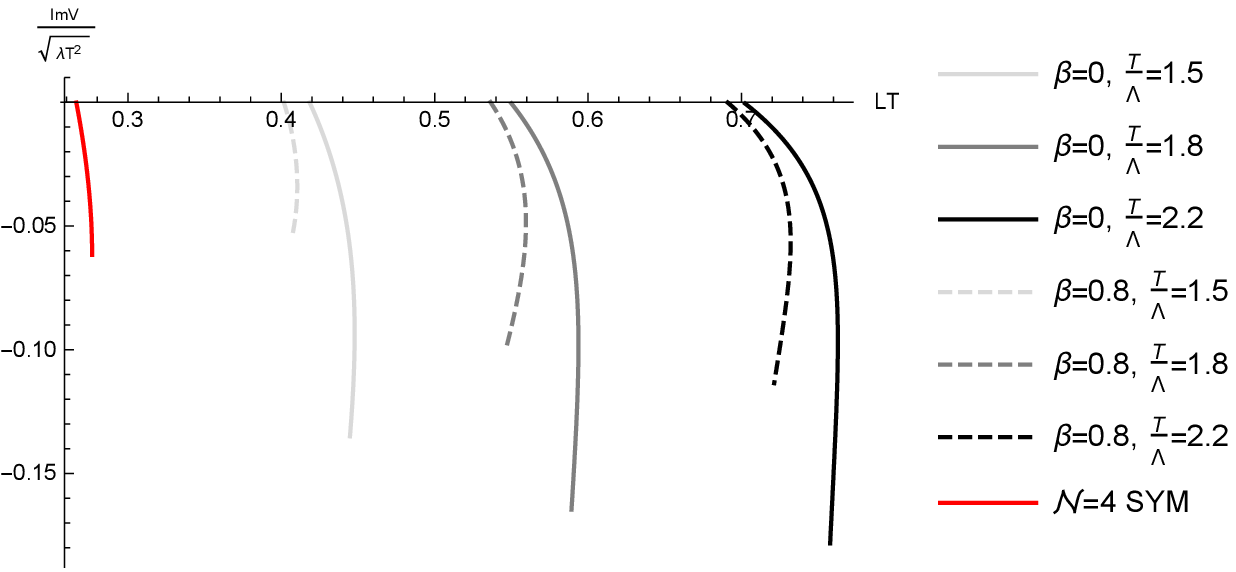}
\end{array}$
\end{center}
\caption{The small black hole. Left: plot of $ LT $ versus $ \xi $ for different values of $ T/\Lambda $ and $ \beta $. Right: $ ImV_{Q\overline{Q}}/\sqrt{\lambda}T $ versus $ LT $ for different values of $ T/\Lambda $ and $ \beta $.}
\label{ggg}
\end{figure}

Similarly with the previous section, we plot $ LT $ and $ ImV_{Q\overline{Q}}/(\sqrt{\lambda}T) $ versus $ \xi $ and $ LT $ respectively, for $ \Theta=0 $. One can see clearly from figures \ref{w2} and \ref{ggg} that the results are similar to the $ \Theta=\pi/2 $: for the large black hole, by increasing $ T/\Lambda $ and $ \beta $, the maximum values of $ LT $ decrease, and the onsets of the $ ImV_{Q\overline{Q}} $ happen for smaller $ LT $s. Therefore, as $ T/\Lambda $ and $ \beta $ increase, the thermal width decreases and the quarkonium dissociates easier. For the small black hole, the values of $ LT_{max} $ increase by increasing $ T/\Lambda $ and decrease by increasing $ \beta $. In the small branch, by increasing $ T/\Lambda $ the suppression will be weaker because the onsets of the $ ImV_{Q\overline{Q}} $ happen for larger $ LT $. Since by increasing $ \beta $ the onsets of the $ ImV_{Q\overline{Q}} $ decrease, the suppression becomes stronger.

To compare the effects of $ T/\Lambda $ and $ \beta $ on $ LT $ and the imaginary potential between $ \Theta=\pi/2 $ and $ \Theta=0 $, we can see clearly that $ LT_{\|} $ is bigger than $ LT_{\bot} $. In addition, we see when the quarkonium is moving transverse to the wind, $ T/\Lambda $ and $ \beta $ affect effectively. In other words, the quarkonium dissociates harder when it moves parallel to the plasma in both large and small black holes. This result is similar to the result of the higher derivative corrections \cite{fad}. Notice that, in the parallel case in the large black hole can be found a moving quarkonium that dissociates herder than in $ \mathcal{N}=4 $ SYM. But, in the small black hole, whether the quarkonium is static or moving, the quarkonium dissociates harder than in $ \mathcal{N}=4 $ SYM.

\section{Entropic force in $ \mathcal{N}=4 $ cSYM}\label{ent}

In this section, following the approach in \cite{hash} we study the effects of $ T/\Lambda $ and $ \beta $ on the entropic force in $ \mathcal{N}=4 $ cSYM. Any fundamental interaction is not explained by the entropic force. The entropic force is an emergent force that relates to the entropy $ S $ and tends to increase its entropy. As proposed in \cite{khar}, the entropic force is,
\begin{equation}
\mathcal{F}=T\dfrac{\partial S}{\partial L},
\end{equation}
where $ T $ is the temperature of the plasma and $ L $ is the inter-quark distance. One can obtain the entropy as,
\begin{equation}
S=-\dfrac{\partial F}{\partial T},\label{sf}
\end{equation}
where $ F $ is the free energy of the quark-antiquark pair. Therefore, in order to evaluate the entropic force, we should calculate $ T $, $ L $, $ F $, and $ S $.
 
From the previous sections, we have the temperature and inter-quark distance as the following,
\begin{equation}
\dfrac{T}{\Lambda}=\dfrac{1-\frac{1}{2}\kappa}{\sqrt{\kappa-\kappa^{2}}},~~~~L=2\int_{r_{c}}^{\infty} dr \sqrt{\dfrac{b(r)a(r_{c})}{a^{2}(r)-a(r)a(r_{c})}},\nonumber\\
\end{equation}
where $ a(r) $, $ b(r) $ and $ a(r_{c}) $ are defined in (\ref{a(r)}), (\ref{b(r)}) and (\ref{ac}) for the case $ \Theta=\pi/2 $ and in (\ref{AR}) for the case $ \Theta=0 $. 

It is obvious from figures \ref{gg}-\ref{ggg}, each $ LT $ plot has a maximum value, called $ c $. If $ LT>c $, the fundamental string breaks into two pieces, and the quarks are screened. In this case, the free energy and entropy are given by,
\begin{equation}
F^{(1)}=\dfrac{1}{\pi\alpha'}\int _{r_{h}}^{\infty} dr,~~~~~~S^{(1)}=\sqrt{\lambda}\theta(L-\dfrac{c}{T}).
\end{equation}
If $ LT<c $, the fundamental string does not break. So, the free energy of the quark-antiquark pair can be obtained as, 
\begin{equation}
F=\dfrac{1}{\pi\alpha'}\int _{r_{c}}^{\infty} dr \sqrt{\dfrac{a(r)b(r)}{a(r)-a(r_{c})}}.
\end{equation}
From (\ref{sf}) we have,
\begin{equation}
S^{(1)}=-\dfrac{\sqrt{\lambda}}{2\pi}\int _{r_{c}}^{\infty} dr \dfrac{\big(a'(r)b(r)+a(r)b'(r)\big) \big( a(r)-a(r_{c})\big) -a(r)b(r)\big( a'(r)-a'(r_{c})\big) }{\sqrt{a(r)b(r)\big( a(r)-a(r_{c})\big)^{3} }},\label{SS}
\end{equation}
where the derivatives are with respect to $ r_{h} $.

\begin{figure}
\begin{center}$
\begin{array}{cc}
\includegraphics[width=73 mm]{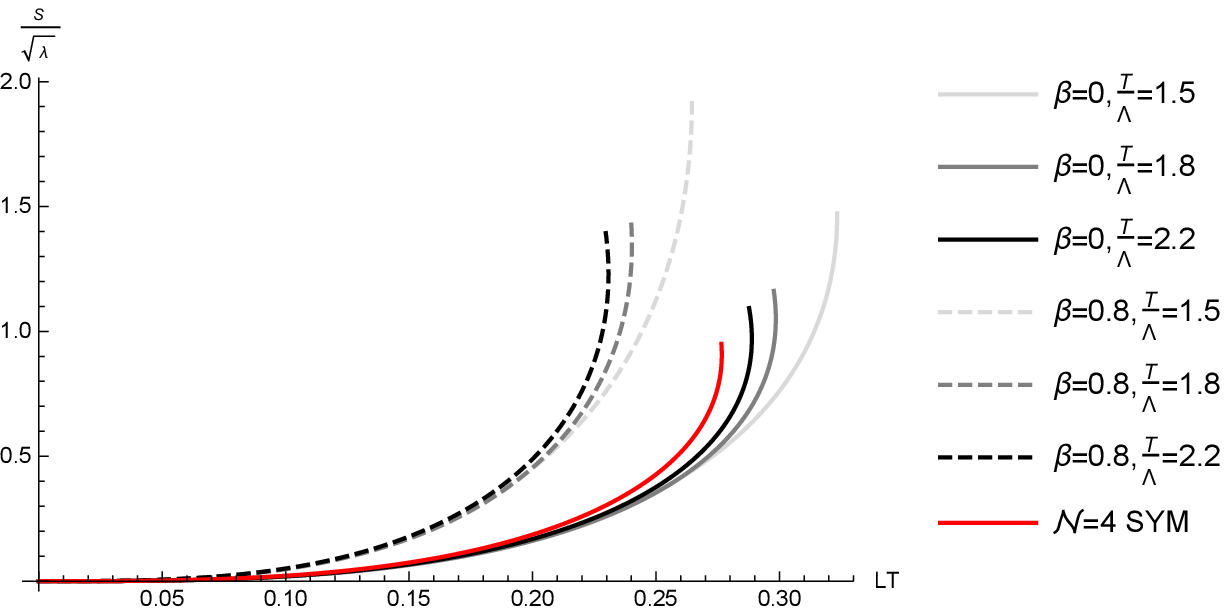}
\hspace{0.2cm}
\includegraphics[width=73 mm]{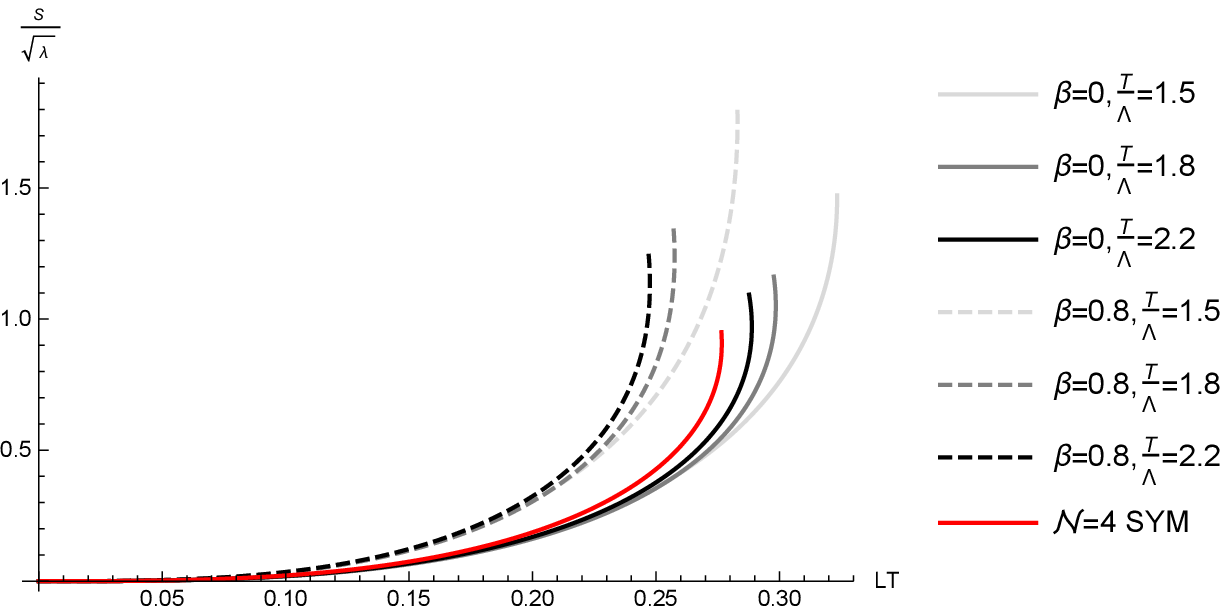}
\end{array}$
\end{center}
\caption{The large black hole. $ S/\sqrt{\lambda} $ versus $ LT $ in different values of $ T/\Lambda $ and $ \beta $: Left: $ \Theta=\pi/2 $, Right: $ \Theta=0 $.}
\label{ent1}
\end{figure}
\begin{figure}
\begin{center}$
\begin{array}{cc}
\includegraphics[width=73 mm]{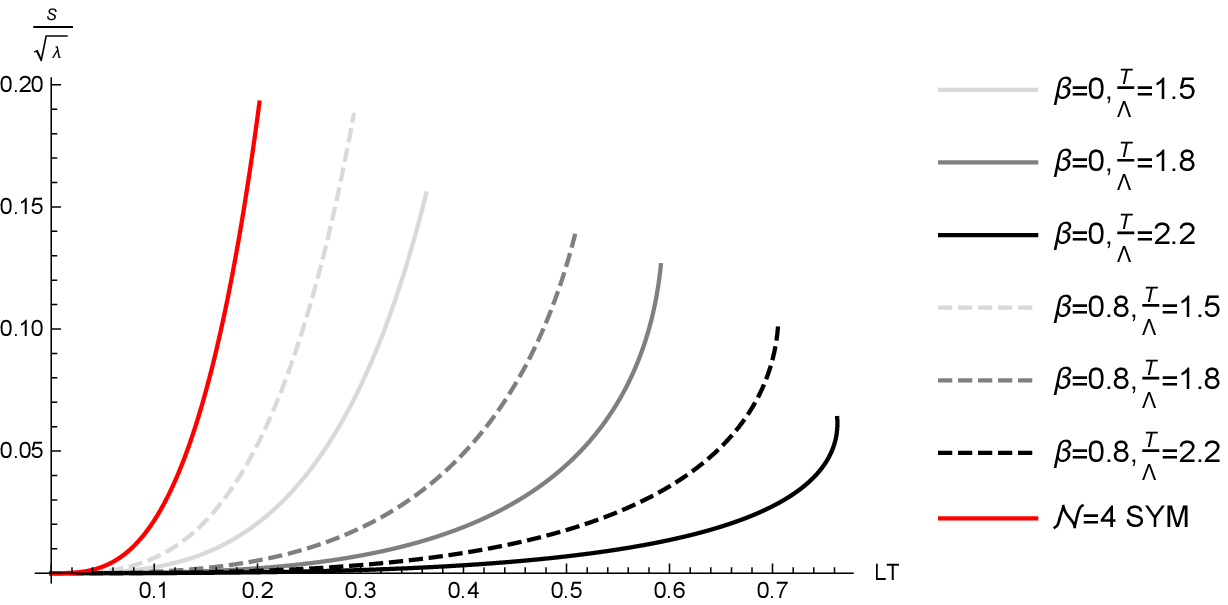}
\hspace{0.2cm}
\includegraphics[width=73 mm]{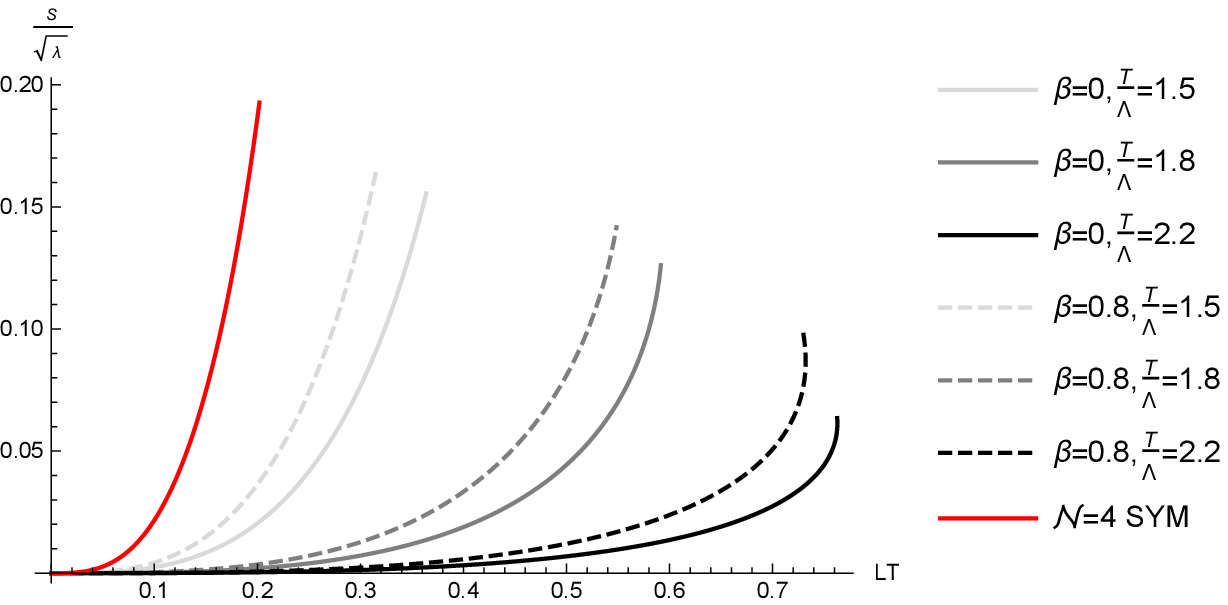}
\end{array}$
\end{center}
\caption{The small black hole. $ S/\sqrt{\lambda} $ versus $ LT $ in different values of $ T/\Lambda $ and $ \beta $: Left: $ \Theta=\pi/2 $, Right: $ \Theta=0 $.}
\label{ent12}
\end{figure}

Now, we analyse the effects of $ T/\Lambda $ and $ \beta $ on the entropic force. For this purpose, we plot $ S/\sqrt{\lambda} $ as a function of $ LT $ numerically for different values of $ T/\Lambda $ and $\beta$ at two cases $ \Theta=\pi/2 $ and $ \Theta=0 $, for the large black hole in figure \ref{ent1} and the small black hole in figure \ref{ent12}. As we see clearly from figure \ref{ent1} for the large black hole, the entropy increases as $ T/\Lambda $ increases. The entropic force in $ \mathcal{N}=4 $ SYM is bigger than in $ \mathcal{N}=4 $ cSYM for the static quarkonium. Also, the entropic force increases by increasing rapidity $ \beta $ so that the entropic force is bigger than in $ \mathcal{N}=4 $ SYM. The result of increasing rapidity leading to increasing the entropic force is consistent with \cite{mov}. 

From figure \ref{ent12} for the small black hole, one can see that increasing $ T/\Lambda $ leads to decreasing the entropic force and increasing $ \beta $ leads to increasing it. Also, the quarkonium dissociates easier in $ \mathcal{N}=4 $ SYM than in the small black hole branch.

Also, from these plots, it is obvious that $ T/\Lambda$ and $ \beta $ have more significant effects on the entropic force when the quarkonium is moving transverse to the wind comparing with the parallel motion which is in agreement with the result of \cite{fad}.

As mentioned before, the entropic force that is in charge of the destruction of the quarkonium is related to the growth of the entropy with the distance. Hence, increasing $ T/\Lambda $ leads to decreasing the thermal width and so the quarkonium dissociates easier in the large black hole. In the small black hole, as $ T/\Lambda $ increases the thermal width increases and hence the quarkonium separates harder. Also, the dissociation quarkonium is easier by increasing $ \beta $ in both the large and small branches.

\section{Conclusion}\label{con}
In heavy-ion collisions, the dissociation of a heavy quarkonium is an important experimental signal for QGP formation. Two present different mechanisms are imaginary potential and entropic force. In this paper, we have investigated the effects of $ T/\Lambda $ and $ \beta $ on these different mechanisms in the rotating black 3-brane solution $ \mathcal{N}=4 $ cSYM. It is shown that the obtained results of these different mechanisms have the same effects. Based on the sing "-" or "+" in (\ref{99}), there are two black hole branches: "-" corresponds to the large black hole and "+" corresponds with the small black hole. 

For the imaginary potential mechanism:\\
In the large black hole branch: we have observed that increasing $ T/\Lambda $ and $ \beta $ lead to decreasing the thermal width and the suppression becomes stronger. In the perpendicular case, the static quarkonium dissociates harder than in $ \mathcal{N}=4 $ SYM but the moving quarkonium dissociates easier. In the parallel case, the static quarkonium separates harder than in $ \mathcal{N}=4 $ SYM and for the moving quarkonium can be found the cases that separate harder. \\
In the small black hole branch: we have obtained that as $ T/\Lambda $ increases the thermal width increases and as $ \beta $ increases it decreases. $ T/\Lambda $ makes the suppression weaker but $ \beta $ has an opposite effect. In this branch, in both perpendicular and parallel cases for the static and moving quarkonium, the quarkonium dissociates harder than in the $ \mathcal{N}=4 $ SYM. Thus, the quarkonium suppression is weaker in the small branch of $ \mathcal{N}=4 $ cSYM than in $ \mathcal{N}=4 $ SYM.

For the entropic force mechanism:\\
In the large black hole branch: we have found that increasing $ T/\Lambda $ and $ \beta $ lead to increasing the entropic force. The entropic force for the moving quarkonium is bigger than in $ \mathcal{N}=4 $ SYM, and the entropic force for $ \mathcal{N}=4 $ SYM is bigger than the static quarkonium in $ \mathcal{N}=4 $ cSYM.\\
In the small black hole branch: the entropic force decreases by increasing $ T/\Lambda $ and increases by increasing $ \beta $. In this branch, the entropic force in $ \mathcal{N}=4 $ SYM is bigger than both static and moving quarkonium in $ \mathcal{N}=4 $ cSYM.

However, it is unknown the connection between these mechanisms yet.

\end{document}